# Electron Attraction Mediated by Coulomb Repulsion

A. Hamo[†,1], A. Benyamini[†,1], I. Shapir[†,1], I. Khivrich[1],

J. Waissman[1,3], K. Kaasbjerg[1,4], Y. Oreg[1], F. von Oppen[2], and S. Ilani[1,*]

[1] *Department of Condensed Matter Physics, Weizmann Institute of Science, Rehovot 76100, Israel.*
[2] *Dahlem Center for Complex Quantum Systems and Fachbereich Physik, Freie Universität Berlin, 14195 Berlin, Germany.*
[3] *current address: Department of Physics, Harvard University, Cambridge, MA 02138.*
[4] *current address: Department of Micro- and Nanotechnology, Technical University of Denmark, DK-2800 Kgs. Lyngby, Denmark .*

[†] These authors contributed equally to this work
[*] Correspondence to: shahal.ilani@weizmann.ac.il

**One of the defining properties of electrons is their mutual Coulombic repulsion. In solids, however, this basic property may change. A famous example is that of superconductors, where coupling to lattice vibrations makes electrons attract each other and leads to the formation of bound pairs. But what if all degrees of freedom are electronic? Is it still possible to make electrons attractive via their repulsion from other electrons? Such a mechanism, termed 'excitonic', was proposed fifty years ago by W. A. Little[1], aiming to achieve stronger and more exotic superconductivity[2–6], yet despite many experimental efforts, direct evidence for such 'excitonic' attraction is still lacking[7]. Here, we demonstrate this unique attraction by constructing, from the bottom up, the fundamental building block[8] of this mechanism. Our experiments are based on quantum devices made from pristine carbon nanotubes, combined with cryogenic precision manipulation. Using this platform we demonstrate that two electrons can be made to attract using an independent electronic system as the binding glue. Owing to its large tunability, our system offers crucial insights into the underlying physics, such as the dependence of the emergent attraction on the underlying repulsion and the origin of the pairing energy. We also demonstrate transport signatures of 'excitonic' pairing. This experimental demonstration of 'excitonic' pairing paves the way for the design of exotic states of matter.**

Soon after the development of the Bardeen-Cooper-Schrieffer (BCS) theory of superconductivity[9], William A. Little proposed a revolutionary idea[1]: Two electrons can attract each other not by a 'phononic' glue, but rather by their repulsion from other electrons. He predicted that if electrons, which are much lighter than ions, form the binding glue, pairing would be dramatically stronger than in conventional superconductors. To test this new form of attraction, Little proposed a concrete realization: a one-dimensional conducting organic chain (Fig. 1a, green) with an array of polarizable sidechains ('polarizers', purple). Each polarizer has a single electron that can hop between two sites – one closer and one further away from the main chain. Due to Coulomb repulsion, an electron traveling down the main chain polarizes the side chains, which in turn would attract another electron in the main chain. Rapidly, this mechanism became popular in attempts to engineer unconventional superconductivity; It was extended to two dimensions[2,3], generalized to attraction at localized sites[10–14] and used in early attempts to explain high Tc superconductivity[15]. It is considered a candidate for the unusual superconductivity[16,17] and pairing[18] observed recently in $SrTiO_3$ interfaces, and it finds analogs in optical systems[19]. Numerous attempts have been made to directly synthesize organic materials having the essential microscopic components[7]. However, to date there is still no direct experimental evidence for an 'excitonic' attraction between electrons.

At the core of Little's proposal is the idea that electrons separate into two groups: some form the 'system' (green, Fig. 1a), where their mutual interaction is to become attractive, while others comprise the surrounding 'medium' (purple, Fig. 1a) that produces the effective glue for the attraction. To make electrons attractive this medium should perform an unusual feat – *flip the sign* of the potential generated by the system electrons, making them look like holes to other system electrons (Fig 1b). This suggests that the medium should effectively have a negative dielectric constant. Little suggested that this can be achieved in the dynamic (retarded) limit, where the system electrons are faster than the medium, thus leaving in their wake a positive polarization cloud that creates an exponentially-weak BCS-like pairing. However, Hirsch and Scalapino showed[5] that the interesting regime of Little's model is in fact in the static (instantaneous) limit, because only then superconductivity dominates over competing orders and exhibits strong Bose-

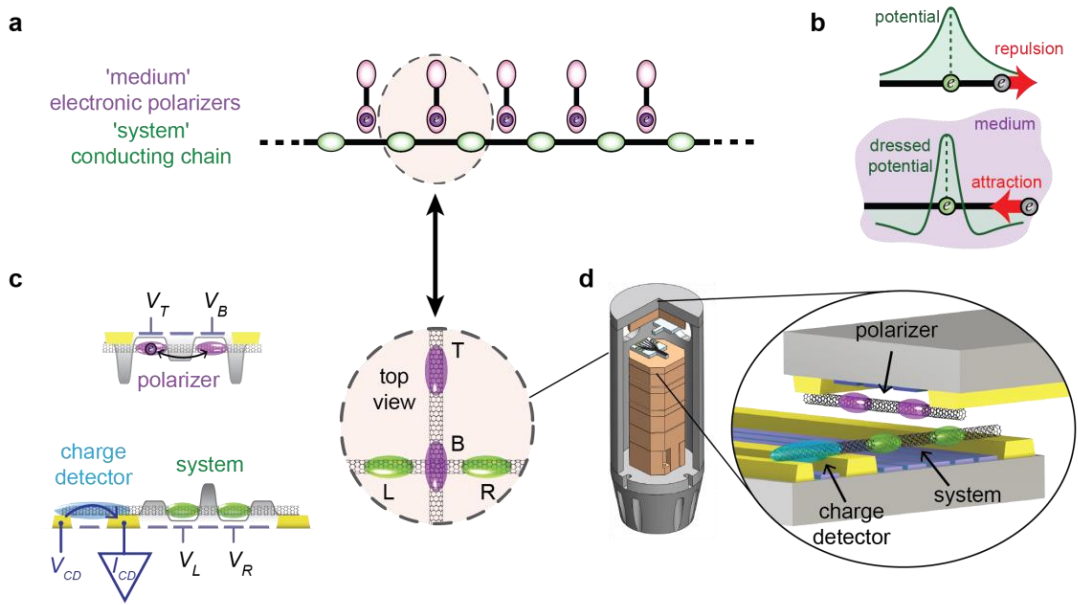

**Figure 1: Model system and experimental realization of its fundamental building block. a,** The organic molecule model system proposed by Little[1], comprising two parts: the 'system' - a one dimensional conducting chain (individual lattice sites marked green), and the 'medium' – an array of side-chain 'polarizers' (purple) each having a single electron that can hop between a site close and a site further away from the chain. The fundamental unit block manifesting attraction is a two-site system and one polarizer (dashed circle). **b,** In a bare electronic system (top) an electron creates a repulsive Coulomb potential (green). Embedding it in a medium that flips the sign of the electrons' potential (bottom) will make this electron attractive to other electrons. **c,** Implementation of the two components, 'polarizer' and 'system', that comprise the fundamental building block. These are fabricated as two separate devices, each having a pristine nanotube (NT) assembled on contacts (yellow) and suspended above an array of gates (blue), which set the potential landscape for the electrons (gray) (see Methods in Supp. Info. S1 for dimensions). The 'polarizer' device (top) has two gate voltages, $V_B$ and $V_T$, that control the potentials of the 'Bottom' and 'Top' dots (purple), and is operated as an isolated dipole, whose sole degree of freedom is that of a single electron hopping between the sites. The 'system' device (bottom) has two gate voltages, $V_L$ and $V_R$, that control the potentials of the 'Right' and 'Left' dots (green). Here the central barrier is opaque and the side barriers are relatively transparent such that electrons enter the two dots from their corresponding leads. An additional dot on a side segment of the same NT (blue) serves as a single electron transistor charge detector: A voltage bias across it, $V_{CD}$, leads to current, $I_{CD}$, that is sensitive to the population of the system dots through weak electrostatic coupling. **d,** A custom built scanning probe microscope (center) operating inside a dilution refrigerator, brings the two oppositely-facing devices into proximity ($\sim 100-150 nm$) such that the polarizer-NT is perpendicular to the system-NT (right) and one of its dots is directly above the system whereas the other is further away (left), creating an analogous structure to that of Little's molecule.

Einstein condensation like pairing. But can a purely electrostatic medium flip the sign of the potential of static electrons and create such strong 'excitonic' attraction?

In this work we demonstrate that 'excitonic' pairing is indeed possible and that the key ingredient is the discreteness of the electrons in the medium. We take a bottom-up approach and construct the minimal building block of Little's model that features this attraction[8] – a two-site system and a single polarizer (dashed circle, Fig 1a) and show that when the polarizer interacts strongly with the system, it can render its electrons attractive.

The 'system' and 'polarizer' are created within two separate carbon nanotubes (NT), each on its own micro-chip. We use our recently-developed[20] nano-assembly technique to suspend each NT between two metallic contacts and above an array of gates (Methods in Supp. Info. S1). In both devices we bias the gates to produce a double-well electrostatic potential along the NT length (Fig. 1c). In the polarizer device, the energy levels in the two wells are placed far from the chemical potential in the leads but close to each other, such that a single electron can hop between the wells but cannot escape to the leads. The polarizer thus operates as an isolated dipole whose sole degree of freedom is the polarization of its electron (Supp. Info. S2). Conversely, in the system device a large central barrier inhibits tunneling between the two wells, but small side barriers allow electrons to enter from their corresponding leads. On a separately contacted side segment of the system NT we create an independent quantum dot that serves as a charge detector, detecting the population of individual system electrons via weak electrostatic coupling. We then mount the two chips, oppositely-facing and perpendicular to each other, in a custom-built scanning probe microscope inside a dilution refrigerator (Fig. 1d). The microscope allows us to control the distance and coupling between the polarizer-NT and the system-NT (inset) and thus to test whether the polarizer alters the behavior of the system electrons in a fundamental way.

We first measure the charge stability diagram of the bare system, without the polarizer. Using the system's left and right gates we scan the potential detuning between the two wells, $\delta V = (V_L - V_R)/2$, and their mean potential, $V = (V_L + V_R)/2$. Changes in the electronic occupation of the system appear as steps in the charge detector current, $I_{CD}$ (Supp. Info. S2). This measurement (Fig. 2a) yields the well-known charge stability diagram of a double-quantum-dot with repulsive electrons (explained in Supp. Info. S3). Tilted charging lines reflect the addition of a single electron to the right or left dot. Near their crossing, the charging lines exhibit a vertical shift (dashed black line) reflecting the fact that if one dot is occupied, the energy to populate the other dot is increased by the Coulombic repulsion between the neighboring electrons. This vertical shift is thus a direct fingerprint of the nearest-neighbor electron repulsion, and its magnitude normalized to energy units yields the strength of this repulsion, $W = 830 \mu eV$.

A fundamentally different behavior is observed when the polarizer is brought into close proximity to the system. Strikingly, we measure in this case a charge stability diagram with an interaction line rotated from vertical to horizontal (Fig. 2b). This rotation implies that the polarizer inverts the interaction between the system electrons from repulsive to attractive. This is best understood by realizing that the ground states at the center of the two charge stability diagrams are fundamentally different: In the bare system, it is a degenerate state between having a single electron on the left or on the right dot (labeled (1,0) and (0,1)). With the polarizer nearby, it becomes a paired ground state, having a degeneracy between the two states with an even electron number, namely (0,0) and (1,1). The odd states, (1,0) and (0,1), become excited states, separated from the ground state by a pairing gap, $\Delta$. This gap is maximal at the center of the horizontal vertex, where its magnitude is given by the length of this vertex ($= 2\Delta$ after normalization, see Supp. Info. S4), reaching $\Delta = 790 \mu eV \sim 8K$.

How can the presence of the medium transform the system electrons from repulsive to attractive? This requires the medium to flip the sign of the electrostatic potential produced by these electrons. The Coulomb potential of a bare electron is positive in its local and neighboring sites, implying on-site and nearest-neighbor repulsion (Fig. 2c). If the on-site repulsion is retained while flipping the potential sign at the nearest neighbor site, the system remains stable, but acquires nearest neighbor attraction. However, any medium based on continuum classical electrostatics can at best screen a potential to zero, but cannot flip its sign (Fig. 2d). What is then the special feature of our polarizer medium that enables sign inversion?

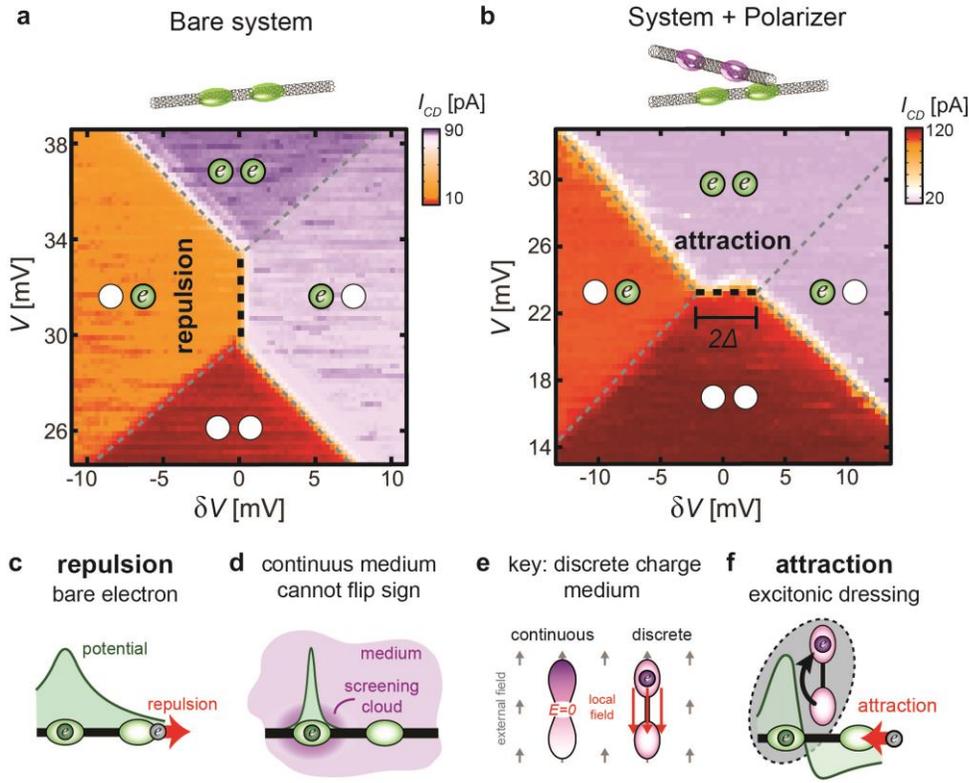

**Figure 2: From repulsive to attractive electrons. a,** Measured charge stability diagram of the bare system: Charge detector current ($I_{CD}$, colormap) plotted as a function of the voltage detuning between its right (R) and left (L) sites, $\delta V = (V_L - V_R)/2$, and the mean gate voltage, $V = (V_L + V_R)/2$. Steps in $I_{CD}$ (dashed gray lines) correspond to single electrons populating the L/R sites (green/white circles label an electron presence/absence). The middle vertical shift (black dashed line) is a direct measure of the Coulomb repulsion between the neighboring electrons (see text). **b,** Similar charge stability diagram, but now with the polarizer-NT positioned nearby (~$125nm$ separation between the NTs). The interaction vertex is now horizontal, reflecting an attraction between the electrons. Along this line (dashed black) the ground state is degenerate between the two even states (0,0) and (1,1) (($n_L, n_R$) represents the number of electrons in the L,R dots), whereas the odd states (1,0) and (0,1) are the excited states, separated by a pairing gap $\Delta$ from the ground state. This gap is maximal at the center of the horizontal line, and its magnitude there is equal to the length of this line (=$2\Delta$, when normalized to energy units, Supp. Info. S4). Note that the charge detector is far away from the right dot and when the polarizer is close their mutual capacitance is strongly screened and thus the charging lines of this dot are harder to observe. **c,** In the bare system, an electron populating the L site generates a Coulomb potential (green) that is positive (repulsive) in both the L and R sites. **d,** Embedding the system in a medium based on continuum electrostatics (purple) can at best screen the potential to zero far away from the electron, but not flip its sign. **e,** The key element for sign inversion is charge discreteness: A single-electron dipole (right) and a similarly-shaped metal (left) will screen an external field differently (gray arrows). In the latter the internal field is nulled whereas in the former it is larger and of opposite sign to the external field (overscreening). This behavior is rooted in the fact that an electron does not repel itself (see text) **f,** With a nearby polarizer, an electron charging the system gets dressed by the polarization (gray ellipse). The electrostatic potential (green) of the dressed particle will be substantially different than that of the bare electron (calculated in Supp. Info. S4). We note that all the measurements in the paper were done with single holes instead of single electrons, but to avoid unnecessary confusion we presented the physics in the language of electrons.

The key element for a sign-inverting medium is charge discreteness[21], and more fundamentally, the fact that a single electron does not repel itself. To understand this, compare the screening by a single-electron dipole to that of a metallic object of similar geometry (Fig. 2e). In a small external field, the charge in the metal will polarize slightly to exactly null the internal field. Conversely, in a dipole with small inter-site tunneling, an entire electron will polarize between the two sites, creating an internal field that is much larger, and of opposite sign to the external field. This forms our basic 'sign-inverter'. Unlike continuous charge in a metal that experiences its own field and thus minimizes the electrostatic energy by nulling the internal field, a single electron does not feel its own field and can thus generate a large over-screening internal field, as long as this minimizes its single-particle energy.

To analyze the sign inversion more microscopically, for the geometry of our system, we describe the polarizer by:

$$H_{pol} = \frac{1}{2}\delta \hat{\sigma}_Z + t\hat{\sigma}_X \tag{1}$$

with $\hat{\sigma}_Z = \psi_T^\dagger \psi_T - \psi_B^\dagger \psi_B$, $\hat{\sigma}_X = \psi_T^\dagger \psi_B + \psi_B^\dagger \psi_T$, where $\psi_T^\dagger$ and $\psi_B^\dagger$ are creation operators of electrons in the two polarizer sites ('top' and 'bottom'), $\delta = \epsilon_T - \epsilon_B$ is the energy detuning between these sites, $t$ is the tunneling amplitude, and we ignored a constant energy offset. The repulsion between system and polarizer electrons leads to a charge-dipole coupling:

$$H_{coup} = \frac{1}{2}U(n_L + n_R)(\hat{I} - \hat{\sigma}_Z), \tag{2}$$

where $n_L, n_R$ are the populations of the left and right system sites, $U$ is the repulsion between an electron in one of these sites and an electron in the bottom site of the polarizer (assuming negligible interaction with the top site), and $\hat{I} = \psi_T^\dagger \psi_T + \psi_B^\dagger \psi_B$. This coupling dresses an electron entering the system by a dipolar excitation of the polarizer (Fig. 2f). In the strong coupling limit[5] ($U > t$) prevailing in Fig. 2b, the electron is dressed by a fully-polarized dipole and the effective potential of the dressed particle is considerably different from that of the bare electron (Supp. Info. S4). In fact, if the charge-dipole coupling exceeds the charge-charge repulsion within the system,

$$H_{repulsion} = W n_L n_R \qquad (3)$$

namely $U > W$, then the dressed potential at the neighboring site has a flipped sign, implying that another electron is attracted to the dressed particle. This attraction reflects the fact that the neighboring electron is more strongly attracted to the vacancy formed in the bottom site of the polarizer than repelled by the original electron that triggered the polarization.

A central prediction of the above model is a direct relation between the emergent attraction and the underlying repulsion. We can test this prediction by changing the separation between the NTs, which modifies the repulsion between system and polarizer electrons ($U$) while keeping the repulsion between the system electrons ($W$) fixed. The measured dependence of $2\Delta$ on $U$ (Supp. Fig. S5) shows a clear linear dependence with unit slope. This result follows from Eq. (1)-(3) in the strong coupling limit (Supp. Info S4) and is a clear demonstration that the observed attraction is driven by repulsion.

What is the optimal detuning of the polarizer for creating the strongest pairing? Since its ability to dress an electron depends on its polarizability, one might expect this to be at zero detuning, where its electronic wavefunction is symmetrically-split between its top and bottom sites and its polarizability maximal. Interestingly, the measured charge stability diagrams at different detunings (Fig. 3a-c) show that the attractive correction to the bare repulsion is not maximal at $\delta = 0$ but rather increases linearly with $\delta$. This observation captures an important point about the origin of the pair binding energy, most easily rationalized by considering a multi-site chain toy model (Fig. 3d,e): Two far-apart electrons on the chain would each be dressed by two adjacent polarizers (Fig. 3d). Conversely, nearest-neighbor electrons share one polarizer and pay one less polarization energy (Fig 3e). This energy gain gives the pair binding energy, which is thus exactly equal to $\delta$, the stored energy in a polarizer, as we observe experimentally. The same simple mechanism of sharing the dressing costs, leading to a pairing energy of magnitude $\delta$ also holds in our two-site case, as we explain in Supp. Info. S6.

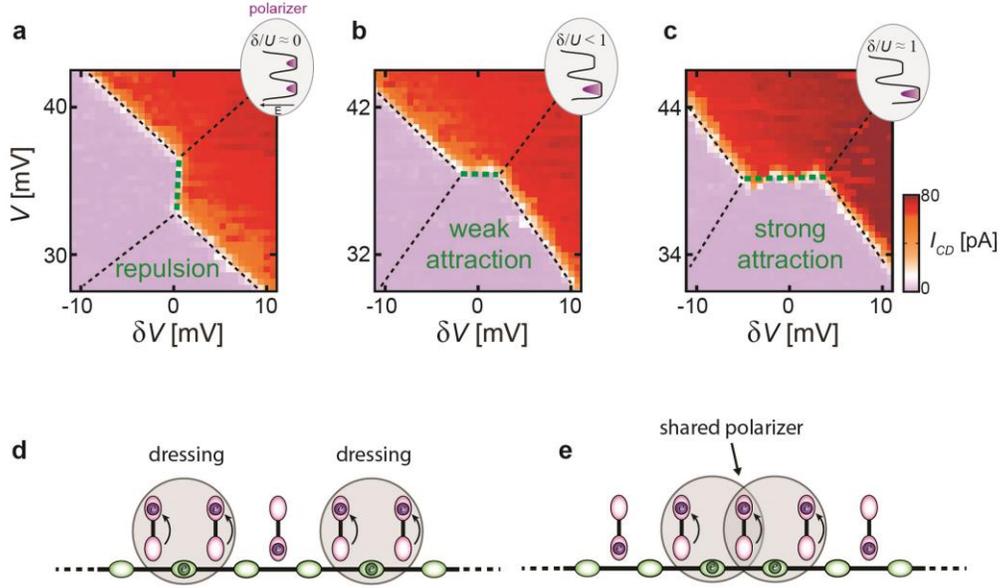

**Figure 3: Pairing energy dependence on the polarizer detuning and the origin of the pair binding energy. a-c,** Charge stability diagrams similar to those in Fig. 2b, measured for different energy detuning of the polarizer, $\delta = 0.39\,meV, 1.01\,meV, 2.57\,meV$. The observed attraction increases linearly with $\delta$ (more data in Supp. Info. S6). Rationalizing this observation using a chain toy model: **d,** two spatially-separated electrons in the chain are dressed each by the polarization of their two adjacent polarizers. **e,** When the electrons are nearest neighbors they share the center polarizer and thus need to polarize one less polarizer. The energy gain, which gives the pairing energy, thus equals $\delta$, as we observe in the experiments in panels a-c. Similar polarizer-sharing argument holds also for the two-site case, as is explained in Supp. Info. S6.

Beyond the ground state stability diagram we can also examine whether pairing is reflected in transport, namely, whether electrons enter the system in pairs. In our setup we can measure the current through the polarizer, and via its strong correlations with the system deduce the behavior of the latter. To simplify the measurement we open the barrier to one of the polarizer dots to strongly couple it to the lead, making the polarizer effectively a single quantum dot coupled to two leads (Fig. 4a). Now, the polarization is between dot and lead rather than two dots. While electrostatically these two cases are quite similar, fundamentally they are very different: whereas the 'excitonic' polarization of an isolated double dot preserves the overall fermion parity of the combined system and polarizer, tunneling of an electron to a lead in the polarizer does not. Yet, establishing pairing also in the latter configuration opens intriguing possibilities to create quantum systems with engineered dissipation, which may even help stabilizing one-dimensional superconductivity[22].

Fig 4b shows the measured polarizer current, $I_{pol}$, with source-drain bias $V_{SD} = -1.3mV$, as a function of its local gate voltage, $V_B$, exhibiting a standard Coulomb blockade peak when the occupation of the polarizer dot is free to fluctuate. Due to the strong coupling to the system, the peak position depends significantly on the electronic occupation of the system (Fig. 4b, various colors). If we fix $V_B$ at $241.5mV$ (dashed line) the polarizer conducts for the odd system occupations (1,0) and (0,1), but is blocked for the even states. The Coulomb peak is pushed well below the Fermi energy for the (0,0) state and well above it for the (1,1) state. This behavior persists throughout the entire charge stability diagram of the system (Fig. 4c, $V_{SD} = 100\mu V$), showing finite current (red) for odd states and negligible current (blue) for even states. Interestingly, the attraction survives even in the dynamic regime with tunneling rates being a significant fraction of the repulsion ($\Gamma_{pol}/U \approx 0.2$, $\Gamma_R/U \approx 0.45$, $\Gamma_L/U \approx 0.23$ where $\Gamma_{L,R,pol}$ are the tunneling rates to the left and right contact of the system and the polarizer contacts respectively). Remarkably, although neither the (0,0) nor the (1,1) states support single-particle current, we observe a finite current peak along their degeneracy line[23,24] (arrow). Theoretical calculations (Fig. 4d) can reproduce this peak only when including correlated many-body events in which an electron hopping in and out of the polarizer is accompanied by a pair of electrons simultaneously cotunneling out of and into the system (Supp. Info. S7).

To better understand the processes underlying the observed current peak, we measure it along a cut perpendicular to the degeneracy line, as a function of $\alpha V$ (where $\alpha$, the independently-measured lever-arm, normalizes $V$ to energy units) and for varying $V_{SD}$ (Fig 4e). In different bias ranges we observe conductance regimes that differ in their turn-on slope, $s = dV_{SD}/d(\alpha V)$, which reflects the number of system electrons participating coherently in the dominant conductance process ($s = 0,1,2$ correspond to processes with zero, one, and two system electrons). At high bias, $V_{SD} > 925\mu eV \approx U$, we observe $s \approx 0$ (black line), showing that a polarizer electron with enough energy to overcome its interaction with the system electrons ($U$) can flow without their cooperation. At intermediate bias, $V_{SD} > 490 \mu eV$, the observed slope is $s \approx 1$ (blue line), reflecting simultaneous cotunneling of one system electron. Below this, and down to zero bias, the slope is $s \approx 2$ (red line), indicating that at low energies a pair of electrons cotunnels into the system in concert with an electron tunneling out of the polarizer. Theoretical calculations (Fig. 4f and Supp. Info. S7) reproduce the main features, although they give a smaller 2e current than observed experimentally, probably because they consider these processes only to the lowest order in tunneling.

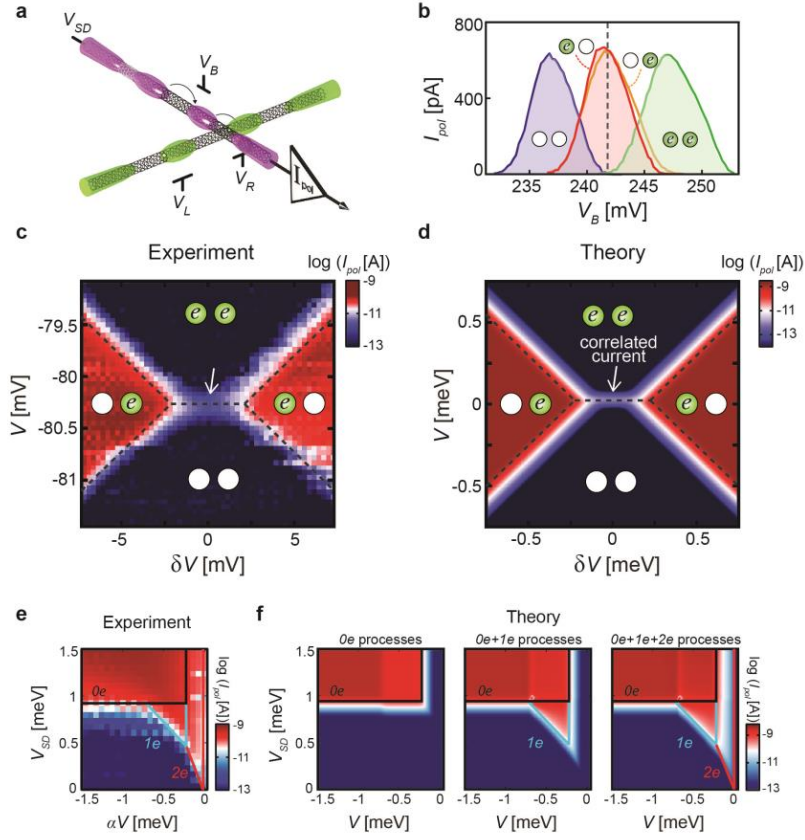

**Figure 4: Transport measurements. a,** Experimental configuration: The polarizer current, $I_{pol}$, is measured with a finite bias on one lead, $V_{SD}$. The barrier between this lead and its nearby dot is reduced such that the latter becomes strongly connected to the lead, effectively making the polarizer-NT a single quantum dot device with two leads. Compared to the experiments in figs. 2,3, in which the polarization occurred internally between two dots, here the polarization is between the dot and its lead. **b,** $I_{pol}$, measured as a function of the local gate voltage, $V_B$, for four different population states of the system: (0,0) blue, (1,0) red, (0,1) orange, (1,1) green. In these measurement the system charge stayed fixed by staying far away from the charging lines of the system. This measurement is performed at a high bias, $V_{SD} = -1.3mV$, which widens the observed charging lines. **c,** $I_{pol}$ (colormap) measured at $V_{SD} = 100\mu V$ as a function of the voltage detuning between the L and R system sites, $\delta V = (V_L - V_R)/2$, and the mean voltage, $V = (V_L + V_R)/2$. Finite current (red) is observed for the odd system states (1,0) and (1,0) and negligible current (blue) is measured for the even states (0,0) and (1,1) apart from a special peak of finite current (white arrow) appearing along their degeneracy line. **d,** A theoretical master equation calculation of $I_{pol}$ for the parameters of the experiment (Supp. Info S7). The theory features a finite current peak on the degeneracy line between the (0,0) and (1,1) state only when considering correlated processes that involve cotunneling of a pair of electrons into the system in concert with an electron tunneling out of the polarizer. **e,** $I_{pol}$ (colormap) measured along a line cutting through the center of the degeneracy line as a function $\alpha V$ and of $V_{SD}$ (where the independently-measured lever-arm, $\alpha = 0.61$, normalizes $V$ to energy units). Three regimes appear at different bias ranges, differing in their turn-on slope, $s = dV_{SD}/d(\alpha V)$, whose value ($s \approx 0$ black line, $s \approx 1$ blue line, $s \approx 2$ red line) reflects the number of system electrons participating in the dominant transport process ($0e, 1e, 2e$) (Supp. Info. S7). **f,** Theoretical master equation calculation of $I_{pol}$ vs. $\alpha V$ and $V_{SD}$ with only $0e$ system process considered (left), with $0e + 1e$ processes (middle) with $0e + 1e + 2e$ processes (left). The $2e$ processes, which involve a pair tunneling in the system, are the dominant process at low bias, as is observed experimentally, although their calculated amplitude is lower than in the experiment, probably because the theory considers them only to the lowest order in tunneling (Supp. Info. S7).

Our work establishes experimentally the fundamental concept of 'excitonic' pairing of electrons, theorized half a century ago. The ability to construct the basic building block of electronic attraction raises many interesting questions regarding the possibility to engineer exotic states of matter by generalizing it to larger systems. Specifically: Is it possible to create an artificial superconductor? What kind of pairing would it have? Would it be stable against competing ground states? (Further discussion in Supp. Info. S8). The beauty of repulsion-driven attraction is that its pairing energy increases linearly with decreasing device dimensions. Our experiments achieved paring energies of $\sim 8K$ with rather large quantum dots ($\sim 400 nm$). By extrapolating this to the $nm$ scale it should be possible to reach energies far in excess of room temperature. Given the tremendous progress in engineering quantum dots in two-dimensional semiconductors[25–28] and down to almost single-atom scale[29], the possibilities to engineer interesting states of matter based on electronic attraction now seem very promising.

**Acknowledgements:** We thank Ehud Altman, Erez Berg, Yuval Gefen, Moshe Goldstein, Ulf Leonhardt, Gil Refael and Amir Yacoby for the stimulating discussions and D. Mahalu for the e-beam writing. K.K. acknowledges support from the Carlsberg Foundation. Y.O. acknowledges support by the Minerva, BSF and ERC Adg grant (FP7/2007-2013 340210). FvO acknowledges support through SPP 1459 and SFB 658. S.I. acknowledges the financial support by the ERC Cog grant (See-1D-Qmatter, No. 647413).

**Author Contributions:** AH, AB, IS and SI performed the experiments, analyzed the data, contributed to its theoretical interpretation and wrote the paper. IS built the scanning probe microscope. IK built custom measurement instrumentation for the experiment. JW designed and fabricated the devices. KK, YO and FvO developed the theoretical model. KK preformed the theoretical simulations.

**Additional Information:** Reprints and permission information is available at http://npg.nature.com/reprintandpermissions. Correspondence and request for materials should be addressed to Shahal Ilani.

# Supplementary Materials

## Electron Attraction Mediated by Coulomb Repulsion

A. Hamo[†1], A. Benyamini[†1], I. Shapir[†1], I. Khivrich[1],

J. Waissman[1], K. Kaasbjerg[1], Y. Oreg[1], F. von Oppen[2] and S. Ilani[1*]

## S1. Methods

**Device fabrication:** Devices were fabricated using our nano-assembling technique, presented in detail in Ref [1]. Specifically, the nanotubes were grown from catalyst using chemical vapor deposition following a standard recipe for single-walled nanotube growth, with argon, hydrogen and ethylene gases. The circuits were patterned on a Si/SiO2 wafer using electron-beam lithography, followed by the evaporation of contacts (10 nm/430 nm Cr/PdAu), gates (4 nm/25 nm Cr/PdAu) and deep reactive ion etching.

**Device properties:** The polarizer device consisted of a nanotube (NT) assembled over a pair of 3.5 $\mu m$-wide contacts and suspended 1.9 $\mu m$ between them at a height of 400$nm$ above 3 gates with 450 $nm$ periodicity. The system device consisted of a NT assembled over a pair of 3.2 $\mu m$-wide contacts and suspended 2.3 $\mu m$ between them at a height of 400$nm$ above 10 gates with 200 $nm$ periodicity, connected in pairs of adjacent gates, effectively yielding 5 gates with 400 $nm$ periodicity. In both devices the size of the quantum dots and their separation were roughly given by the periodicity of the gates and were ~400 − 500 $nm$. In the system device we used an additional side segment of the same NT as the charge detector. This segment was suspended 700 $nm$ between one of the main contacts and a third contact (see Fig. 1c bottom in the main text) and gated by the silicon back gate. In the transport measurements of Fig. 4 we reversed the roles of the two devices, to achieve better control over the tunneling barriers in the polarizer. The overall contact resistance in both devices was on the order of ~500 $k\Omega$.

**Scanning probe microscopy:** The relative positioning of the two NT devices was done using a home built microscope, based on Attocube coarse and fine piezo positioners. The two devices were connected electrically through 14 coaxial lines to the device mounted on the piezo motors and 20 lines to the second device, which is fixed in space. The microscope was mounted inside a dilution fridge running at a base temperature of 12mK. The electron temperature was determined to be ~100$mK$ in both devices based on the measured width of their corresponding Coulomb blockade peaks. Due to small angles between the planes of the two samples, the distance between the two NTs was limited to ~100$nm$ .

**Correcting for cross-capacitance between opposite devices**: When the system and polarizer are at close proximity, the gate electrodes of one device are capacitively coupled also to the other device (Fig. S1a) and thus affect the potential of the latter. We correct for these cross couplings by working in a 'rotated' gate voltage basis, which uses the proper linear combination of gate voltages such that each gate voltage in the rotated basis controls only the potential of a single dot. To demonstrate this with a specific example we plot in Fig. S1b the measured charge detector current, $I_{CD}$, as a function of the two non-rotated gate voltages: $\tilde{V}_L$, the voltage on the gate beneath the 'left' dot in the system NT and $\tilde{V}_B$, the voltage on the gate above the 'bottom' dot in the polarizer NT. The charging of an electron into the 'left' dot is visible as a step in the value of $I_{CD}$ (marked by a dashed black line). Notably, this charging line is slanted in $\tilde{V}_L$-$\tilde{V}_B$ plane, indicating that both gates are capacitively coupled to this dot. To negate the effect of the 'bottom' gate on the left dot we transform to a rotated basis, $V_L = \tilde{V}_L$ and $V_B = \tilde{V}_B + s \cdot \tilde{V}_L$ , with a properly chosen coefficient $s$. As can be seen in Fig. S1c after this transformation the 'left' dot is effected only by $V_L$ and not by $V_B$. In a similar manner we measure the cross-capacitance between all system and polarizer gates to any of the dots in both these devices, and using similar transformations obtain a rotated basis in which a certain gate voltage affects only a certain dot. These rotated gate voltages, $V_L$, $V_R$, $V_B$ and $V_T$ are the ones used throughout the paper and in the other sections in the Supplementary Information.

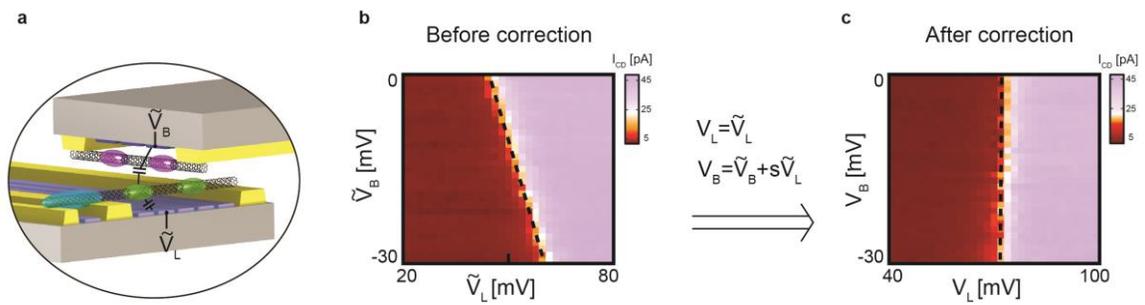

**Figure S1: Correcting cross-capacitance effects between opposite devices. a.** An illustration of the system and polarizer, showing the capacitance between the left gate of the system and the bottom gate of the polarizer to the left dot in the system NT. **b.** A color map of the current through the charge detector (blue in panel a) as a function of the uncorrected voltages $\tilde{V}_L$, $\tilde{V}_B$. Charging of an electron in the left dot appears as a sharp

change in the current along a slanted line (dashed black). **c.** The same measurement using the corrected voltages $V_L$, $V_B$.

## S2. Operation point of the polarizer and system devices

The polarizer device in our experiments comprises two potential wells, which we term the top (T) and the bottom (B) wells. These wells are defined using three barriers: two naturally-formed barriers near the contacts and a central barrier whose height is controlled by a voltage on a center gate, $V_C$ (Fig. S2a). The potential of the two wells is controlled by the voltages on their corresponding plunger gates, $V_B$ and $V_T$. The measured charge stability diagram of the polarizer device is shown in Fig. S2b. Note that all the measurements in this paper were done with holes (in the nanotube's valence band) rather than electrons (in the nanotube's conduction band). The entirety of the physics described in this paper is completely invariant to this flip, however, if this physics was described in the language of holes this might have cause an unnecessary confusion, especially since in Little's model we often need to think of holes as the vacancies that the electron leaves. To avoid such a confusion we present the physics everywhere in the paper in the language of electrons. Namely, we consider the absence of a hole in the dot as the presence of an electron.

For the measurement in Fig. S2b, we configure the system-NT as a charge detector by defining a single quantum dot along the entire length of its central suspended segment whose conductance, $G$, changes when electrons populate the polarizer. The figure plots $G$ as a function of $V_B$ and $V_T$, exhibiting steps (marked by dashed white lines) that correspond to the charging of the 'top' and 'bottom' dots.

In the experiments described in Fig. 2 and Fig. 3 of the main text we operate the polarizer nanotube device as an isolated dipole, whose sole degree of freedom is that of a single electron that can hop between the two sites, but cannot exit to the leads. This is achieved by tuning the energy levels of the two wells away from the chemical potential in the leads, but close to each other, such that the only allowed transition is that of a single electron between the two sites. Specifically, in the experiments presented in these figures we use the (2,3) ↔ (3,2) transition, having two frozen charges in each dot and one additional charge that can move between them. This charge forms the dipolar degree of freedom, and in this subspace the polarizer is described by a two level isospin basis, $|⇓⟩$ and $|⇑⟩$ corresponding to the polarization of an additional electron to the T or B sites respectively. We note that the frozen charges in the polarizer do not affect the physics of attraction by repulsion, and that we have observed similar attraction also when the polarizer was operated around the $(0,1) \leftrightarrow (1,0)$ transition.

A more detailed charge stability diagram of the vertex around this transition is shown in Fig. S2c. Here we plot the derivative of the charge detector current with respect to the bottom-well gate in the polarizer device, $dI/dV_B$. The charging lines that correspond to charges entering the two polarizer wells appear as positive peaks in $dI/dV_B$ (marked with tilted dashed white lines as a guide to the eye). At the center of the diagram we observe a dip in $dI/dV_B$ (blue, marked by a vertical dashed white line), which corresponds to the internal transition of an electron between the two wells, namely, to a transition between the two polarizer states, $|⇑\rangle$ and $|⇓\rangle$ (labeled). In the experiment, we position ourselves at the center of this line and move perpendicular to it by varying the detuning, $\delta$ (horizontal dashed black line).

We can also extract the tunneling element between the two polarizer dots, $t$, from a line-cut measurement done along the detuning axis, $\delta$, as is shown in the inset to Fig. S2c. In this measurement we plot the transconductance, $dI/dV_B$, normalized by the gain of the charge detector, $dI/dV_g$ (where $V_g$ is applied to all the gates of the charge detector together). The normalization cancels out small changes in the gain of the charge detector induced by the polarizing electron[2]. This measurement reflect the polarizability of the polarizer, peaking at zero detuning. The width of the observed peak gives directly the inter-dot tunneling, which for the measurement in the inset of Fig. S2c equals $t \sim 160 \mu V$. For the same conditions, the coupling $U$ between the polarizer bottom dot and the system dots is on the order of $2\ mV$, placing our measurements in the strong coupling region $U > t$.

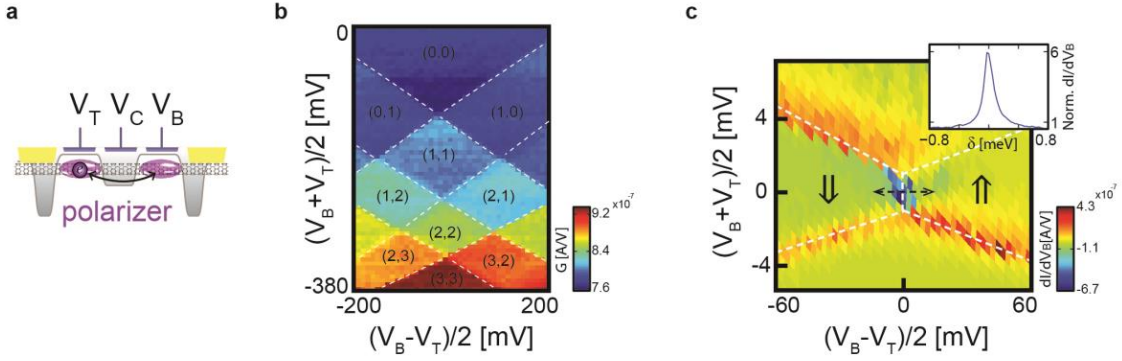

**Figure S2: Polarizer device charge stability diagram. a,** Illustration of the polarizer device. **b.** The full polarizer charge stability diagram. For this measurement we use the system-NT as a charge detector, by forming a single quantum dot along the entire length of its central suspended NT segment. We measure the differential conductance through the dot, $G = dI/dV$, as a function of the gate detuning, $(V_B - V_T)/2$, and their mean, $(V_B + V_T)/2$. The charge state of the top and bottom dots is labeled as $(n_T, n_B)$. Transitions between charge states are marked with dashed lines as guide to the eye. **c,** The polarizer stability diagram measured around the $(2,3) - (3,2)$ vertex used for the experiments. Here the colormap plots the derivative of the charge detector current with respect to the one of the bottom-gate voltage in the polarizer, $dI/dV_B$, measured as a function of the voltage detuning between the gates of the top and bottom dots, $(V_B - V_T)/2$, and their mean, $(V_B + V_T)/2$. Positive peaks in $dI/dV_B$ (red) correspond to charging lines of the individual dots (marked by tilted dashed white lines). Negative peak in $dI/dV_B$ (blue, marked by a vertical dashed white line) reflects the transition of a single electron between the wells, namely, to the transition between the two polarization states of this polarizer, labeled as ⇑ and ⇓. In the experiment, we work near the center of this line and move perpendicular to it along the dashed black detuning line. Inset: Plot of the transconductance, $dI/dV_B$, normalized by the gain of the detector, $dI/dV_g$ (where $V_g$ is applied to all the gates of the charge detector), as a function of the detuning normalized to energy units, $\delta$, along the dashed black line.

The system NT device has also two potential wells, termed left (L) and right (R), whose potential we control using two plunger gates, $V_L$ and $V_R$. The dots are confined by three gate-induced barriers, one between the dots and two between each dot and its corresponding lead (Fig. S3a). To measure the charge stability diagram of the system, we use the local charge detector which is located on a separately contacted side segment of the same NT (blue, Fig. S3a). Fig. S3b, shows the transconductance $dI_{CD}/dV_L$, measured as a function of $V_L$ and $V_R$, exhibiting the charging lines of electrons in the system NT. Since the charge detector is located closer to the left dot, the transconductance signal is much stronger for charge transitions in the left dot as compared to those in the right dot (marked by dashed lines). Consequently, not all of the transitions in the left dot are directly visible, however, they can be clearly observed as kinks in the charging lines of the right dot, formed by the repulsion between the electrons in the two dots. The experiments in Figs. 2,3 of the main text are done along the vertex with a single charge in each dot, namely, the $(0,0) - (1,0) - (0,1) - (1,1)$ vertex.

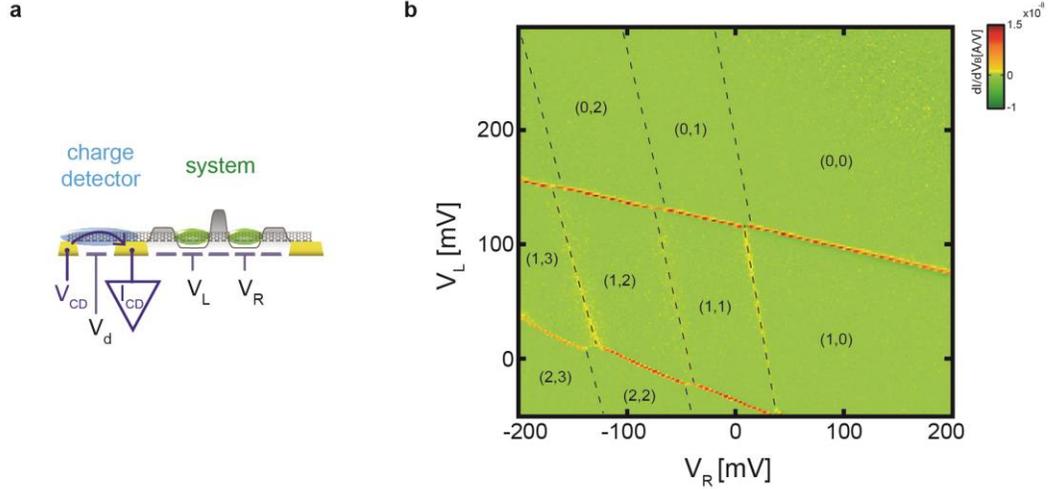

**Figure S3: The overall charge stability diagram of the system NT. a.** Illustration of the system NT device. The 'left' and 'right' dots (green) are separated by three gate-defined barriers. The potential of the two wells is determined by the left and right gate voltages, $V_L$ and $V_R$. An additional dot formed on a side segment of the same NT (blue) acts as an independent charge detector, and can be tuned with gate voltage $V_d$. **b.** The charge stability diagram of the system, measured using the charge detector. In colormap we plot the derivative of the charge detector current with respect to the voltage of the left gate, $dI_{CD}/dV_L$, measured as a function of $V_L$, and $V_R$. The charge carrier populations of the left and right dots are labeled as $(n_L, n_R)$. Charging lines of the right dot, which are weaker or absent due to the larger distance between this dot and the charge detector, are marked with dashed lines.

The charge detector we use in our experiments is made of an independent quantum dot, which is located on a side segment of the same nanotube we use to create the system (blue, Fig. S3a). This dot has weak electrostatic coupling to the system (interaction energy of $\sim 70\ \mu eV$), which is an order of magnitude smaller than all other interaction energies between the dots in the system and polarizer. Thus it has no significant effect on their physics. Despite the weakness of the coupling, we are able to use this charge detector to detect charging events in the system. To do this we tune it to a sensitive point on its Coulomb blockade (CB) peak, where its current has a large derivative with respect to its local gate, $V_d$ (Fig. S4a). When the charge configuration of the system changes, the CB peak shifts and we detect a step in the current. The current can either increase or decrease with increasing charge in the system, depending on weather we are on the left side of the CB peak where the gain $dI_{CD}/dV_d$ is positive, or on the right side where it is negative. The workpoint of the detector has no effect on the physics in the system, as can be seen from two different measurements of the system charge stability diagram, taken at two different sides of the detector's CB peak (Fig. S4b,c).

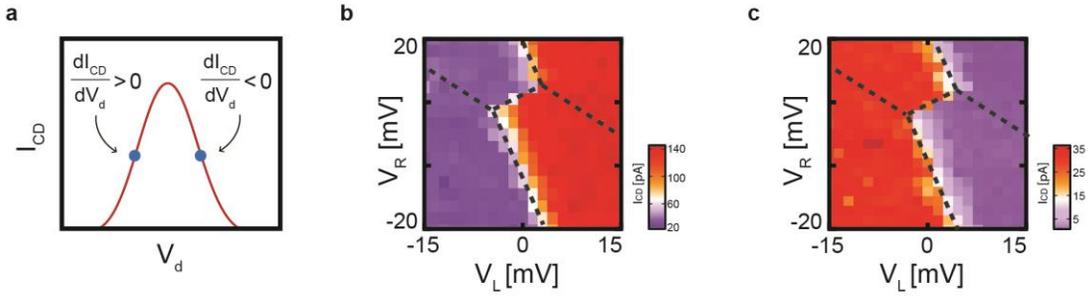

**Figure S4: Measurement of the system using the charge detector at different workpoints. a.** Illustration of the CB peak of the charge detector as a function of its local gate voltage $V_d$. **b.** The charge stability diagram of the system, measured using the charge detector set at a workpoint with positive gain (left blue dot in a). **b.** A similar measurement taken with the charge detector set at a workpoint with negative gain (right blue dot in a). The different workpoint of the charge detector has no effect on the physics, as can be seen from the same charge stability diagram obtained in both measurements.

## S3. The charge stability diagram of a double quantum dot

The charge stability diagram of a double quantum dot was used in the main text as a tool to identify the interaction between electrons on the two neighboring sites. Here we give a brief introduction to these stability diagrams, and explain the voltage axes used in the measured diagrams in the paper.

The charge stability diagram of a double quantum dot plots how the occupation of two coupled dots ('left' and 'right') depends on their local gate voltages, $V_L$ and $V_R$. Fig. S5a illustrates the generic diagram observed in double dots[3], showing the ground state population of the left and right dots, labeled $(n_L, n_R)$, as a function of $V_L$ and $V_R$. Changing $V_L$ (/$V_R$) changes the energy of an electronic level in the left (/right) dot. When this level crosses the chemical potential in the leads it gets populated by a single electron. In the diagram in Fig. S5a the population of the left dot occurs along the vertical blue lines and that of the right dot along the horizontal blue lines. Repulsive interactions between the electrons in the two dots cause the lines to split near their crossing point along a diagonal line (red). This is because the energy to charge an electron in one of the dots increases when it's neighboring dot is populated by an amount given by the repulsion energy between the two. In the paper we present the charge stability diagram in a rotated voltage coordinate system, given by the mean voltage, $V = (V_L + V_R)/2$, and the detuning voltage, $\delta V = (V_L - V_R)/2$ (Fig. S5b). In this representation the repulsive interaction line at the center of the vertex is vertical. As discussed in the main text, the interaction line for attractive electrons is horizontal.

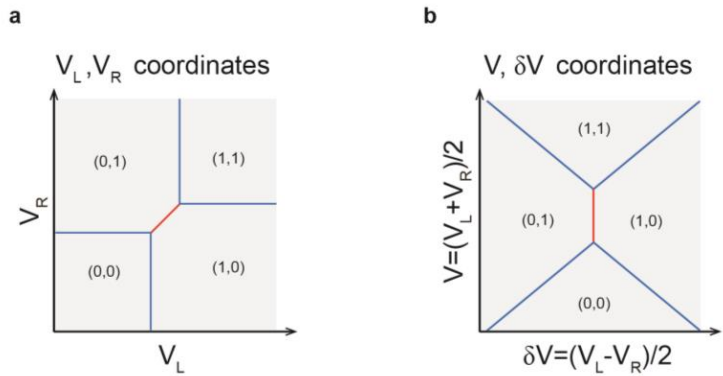

**Figure S5: Charge stability diagram in different coordinate systems. a.** Charge stability diagram in the $V_L, V_R$ coordinate system. The charging lines are shown in blue, and the interaction line in red. The charge state is denoted by $(n_L, n_R)$ **b.** The same diagram in the $V, \delta V$ coordinate system. The repulsion line (red) in these coordinates is vertical, while an attraction line would be horizontal.

## S4. Derivation of the effective low energy attraction Hamiltonian

In this section we derive the effective low-energy Hamiltonian of our system by projecting out the high energy states of the polarizer. A similar derivation was originally performed by Hirsch and Scalapino[4] for the one-dimensional chain model by Little[5], systematically covering the entire range of parameter space. In this section we re-derive their results only for the special case of two lattice sites and a single polarizer, in the strong coupling regime and static limit, which are relevant for understanding the experiments presented in Fig. 2 and 3 of the main paper. The dynamic limit relevant for Fig. 4 of the main paper is analyzed separately in section S5 below.

**Deriving the effective low-energy Hamiltonian**

If we add to equations (1)-(3) in the main text the single particle energy terms of the system, we obtain the full Hamiltonian of the combined system + polarizer, which reads:

(S4.1)
$$H = H_{sys} + H_{pol} + H_{coup}$$
$$H_{sys} = \mu_L n_L + \mu_R n_R + W n_L n_R$$
$$H_{pol} = \frac{1}{2}\delta \hat{\sigma}_Z + t\hat{\sigma}_X + \bar{\epsilon}\hat{I}$$
$$H_{coup} = \frac{1}{2}U(n_L + n_R)(\hat{I} - \hat{\sigma}_Z),$$

where in the system Hamiltonian $\mu_L = \epsilon_L - v_L$, $\mu_R = \epsilon_R - v_R$ are the chemical potentials of the left and right dots, $\epsilon_L$, $\epsilon_R$ are their single particle energies, $v_L$, $v_R$ are their gate potentials in units of energy, $n_L$, $n_R$ are their occupations, and $W$ is the nearest-neighbor repulsion. In the polarizer Hamiltonian $\hat{\sigma}_Z = \psi_T^\dagger \psi_T - \psi_B^\dagger \psi_B$, $\hat{\sigma}_X = \psi_T^\dagger \psi_B + \psi_B^\dagger \psi_T$, $\hat{I} = \psi_T^\dagger \psi_T + \psi_B^\dagger \psi_B$, where $\psi_T^\dagger$, $\psi_B^\dagger$ are the creation operators of an electron in the top and bottom sites, $t$ is the hopping amplitude between these sites, and $\delta = \epsilon_T - \epsilon_B$ and $\bar{\epsilon} = (\epsilon_T + \epsilon_B)/2$ are the detuning and mean energies of these sites. This description of a polarizer double-dot via spin-like operators is a natural consequence of the fact that we are confined to the subspace of a single electron occupation in the polarizer. The $\bar{\epsilon}\hat{I}$ term in this Hamiltonian is a constant and was therefore ignored in the main text. In the coupling Hamiltonian, the charge-dipole coupling element, $U$, represents the repulsion between an electron in the left or right sites of the system and an electron in the bottom site of the polarizer, and we have assumed that the top site is far away so its interaction with the system is negligible.

Since we do not consider tunneling between the system sites, $n_L$ and $n_R$ are good quantum numbers. Diagonalizing the full Hamiltonian we obtain the eigenvalues in $n_L, n_R$ for each of the polarizer states:

(S4.2) $\quad E^{\pm}_{n_R,n_L} = \mu_L n_L + \mu_R n_R + W n_L n_R + \bar{\epsilon} + \frac{1}{2}U(n_L + n_R) \pm \sqrt{\left(\frac{\delta - U(n_L + n_R)}{2}\right)^2 + t^2}$

For each configuration of the system we project out the polarizer configuration that leads to the high energy state. The resulting low-energy sector can be described by the following effective system Hamiltonian:

(S4.3) $\quad H_{eff} = \tilde{\mu}(n_L + n_R) + \widetilde{\delta\mu}(n_L - n_R) + \tilde{U} n_L n_R + E_{00}$

The parameters of this Hamiltonian, determined from the $E^{-}_{n_R,n_L}$ found above, are:

(S4.4) $\quad E_{00} = \bar{\epsilon} - \sqrt{\frac{1}{4}\delta^2 + t^2}$

(S4.5) $\quad \tilde{\mu} = \frac{\mu_L + \mu_R}{2} + \frac{U}{2} - \sqrt{\frac{1}{4}(\delta - U)^2 + t^2} + \sqrt{\frac{1}{4}\delta^2 + t^2}$

(S4.6) $\quad \widetilde{\delta\mu} = (\mu_L - \mu_R)/2$

(S4.7) $\quad \tilde{U} = W - \sqrt{\frac{1}{4}(\delta - 2U)^2 + t^2} + 2\sqrt{\frac{1}{4}(\delta - U)^2 + t^2} - \sqrt{\frac{1}{4}\delta^2 + t^2}$

The quadratic term in this Hamiltonian, $\tilde{U} n_L n_R$, captures the effective interaction between the electrons in the system. The first term in $\tilde{U}$ (eq. S4.7) is the repulsive potential that a bare electron (without a polarizer) produces in its nearest neighbor site. The next three terms add a negative (attractive) contribution that is due to the dressing of this electron by the polarizer. For a large enough $U$ these attractive terms become larger than the bare repulsion, $W$, resulting in an attractive interaction between the two electrons.

When the two electrons attract each other, the two even states (0,0) and (1,1) can become degenerate. This happens at a certain chemical potential, which can be determined by requiring $E^{-}_{00} = E^{-}_{11}$ in eqs. (S4.2)-(S4.7), giving $\tilde{\mu} = -\frac{1}{2}\tilde{U}$. The two odd states (1,0) and (0,1) then become the excited states, separated by a pairing gap from this ground state. This gap is determined by smallest excitation energy:

(S4.8) $\Delta = \min\{(E_{10}^- - E_{00}^-), (E_{01}^- - E_{00}^-)\} = \min\{-\frac{1}{2}\widetilde{U} + \widetilde{\delta\mu}, -\frac{1}{2}\widetilde{U} - \widetilde{\delta\mu}\} = -\frac{1}{2}\widetilde{U} - |\widetilde{\delta\mu}|$

Giving:

(S4.9) $2\Delta = \sqrt{\frac{1}{4}(\delta - 2U)^2 + t^2} - 2\sqrt{\frac{1}{4}(\delta - U)^2 + t^2} + \sqrt{\frac{1}{4}\delta^2 + t^2} - W - 2|\widetilde{\delta\mu}|$

The largest gap is obtained when the chemical potentials of the left and right dots are degenerate, $\widetilde{\delta\mu} = 0$, which occurs at the center of the charge stability diagrams in Figs. 2-4 in the main text.

### The strong coupling limit ($U \gg t$)

To derive the formula for the pairing gap in the strong coupling limit ($U \gg t$) we will take the $t \to 0$ limit. This limit will provide an accurate description of the gap apart from $\sim O(t)$ corrections near the points $\delta = 0, U, 2U$. In this limit, Eq. (S4.9) simplifies to:

(S4.10) $2\Delta = \frac{1}{2}|\delta - 2U| - |\delta - U| + \frac{1}{2}|\delta| - W - 2|\widetilde{\delta\mu}|$

For small detunings, $\delta < U$, and left-right symmetric system, $\widetilde{\delta\mu} = 0$, we get that the pairing energy grows linearly with $\delta$:

(S4.11) $\boxed{2\Delta = \delta - W \quad \text{for } \delta < U}$

capturing the detuning dependence that we measured in Fig. 3a-c in the main text and Fig. S7 below.

The maximal gap is obtained when $\delta = U$:

(S4.12) $\boxed{2\Delta_{\max} = U - W \quad \text{for } \delta = U}$

fitting well the measured height dependence in Fig. S6 below.

### Extracting the pairing energy from the length of the vertex in the charge stability diagram

In the main text we extracted the magnitude of the pairing gap from the length of the horizontal interaction vertex in the charge stability diagram (e.g. in Fig 2b). In this section we derive this relation formally in the strong Coupling limit.

Rewriting eq. (S4.10) as:

(S4.13) $2\Delta = 2\Delta_{\widetilde{\delta\mu}=0} - 2|\widetilde{\delta\mu}|$

we see that the gap is maximal for $\widetilde{\delta\mu} = 0$ (center of the stability diagram) and decreases linearly with $\widetilde{\delta\mu}$, until it becomes zero at the left and right edges of the horizontal vertex, namely $\widetilde{\delta\mu}_{right\ edge} = -\widetilde{\delta\mu}_{left\ edge} = \Delta_{\widetilde{\delta\mu}=0}$. Since $\mu_L = \epsilon_L - v_L$, $\mu_R = \epsilon_R - v_R$ and the $\epsilon_L, \epsilon_R$ are constants then we get that the size of the horizontal vertex in terms of $\delta v = (v_L - v_R)/2$ is:

(S4.14) $\quad L_{vertex} = \delta v_{left\ edge} - \delta v_{right\ edge} = \widetilde{\delta\mu}_{right\ edge} - \widetilde{\delta\mu}_{left\ edge} = 2\Delta_{\widetilde{\delta\mu}=0}$

## S5. Dependence of pairing energy on the separation between the system-NT and the polarizer-NT

By changing the separation between the system-NT and polarizer-NT, using the scanning probe microscope, we can directly control one of the important parameters in the model: the charge-dipole coupling between the system electrons and the polarizer's polarization (Eq. 2 in the main text). Microscopically, this coupling is given by the repulsion between an electron in left or right sites of the system and electron in the bottom site of the polarizer, $U$ (where we assumed negligible interaction with the top polarizer site, which is a valid assumption for the geometry of our experiment). By changing the NT separation we can thus directly change the repulsion-driven term in the Hamiltonian, $U$, and check how it affects the emergent attraction between the system electrons, as reflected by their pairing energy, $\Delta$. Fig. S6 shows the measured dependence of $2\Delta$ on $U$, where for each point in this graph we have determined $U$ from a measurement of the repulsive interaction vertex between an electron in the system and that on the bottom site of the polarizer, and $2\Delta$ from the size of the attractive vertex in an independent measurement of the system's charge stability diagram, similar to that shown in Fig. 2b in the main text. In addition, at each point $\delta$ was tunned to $\delta \approx U$, which gives the maximal pairing (see eq. S4.12). Notably, for small values of $U$ the system electrons are repulsive ($2\Delta < 0$), but with increasing $U$ this repulsion turns into attraction ($2\Delta > 0$). The data is fitted reasonably well by a line with unit slope, $2\Delta = U - W$ (dashed blue), which is the predicted dependence in the strong coupling limit (eq. S4.12). This dependence is thus a clear indication that the observed attraction is driven by repulsion.

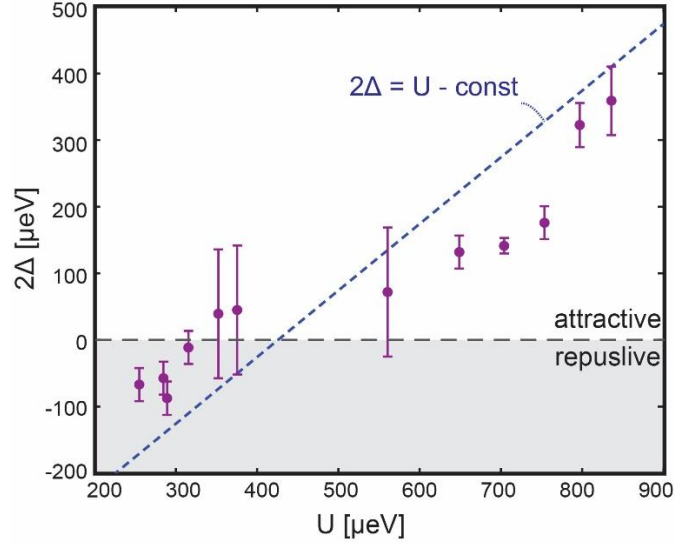

**Figure S6: Dependence of attraction on repulsion.** We change the strength of the repulsion between electrons in the system and those in the polarizer by changing the separation between the polarizer-NT and system-NT and measure how this affects the magnitude of the emergent attraction between electrons in the system. The different points in the graph are measured at different NT-NT separations ($\sim 150 nm - 1800 nm$). For each point we determine the strength of the repulsion, $U$, from a measurement of the repulsive charge stability diagram between an electron in the system and an electron in the bottom site of the polarizer. For the same point we extract $2\Delta$ from an independently-measured charge stability diagram of the two electrons in the system, similar to that in Fig. 2b in the main text, where $2\Delta$ is given by the length of the horizontal vertex. Negative values of $2\Delta$ correspond to the length of a vertical (repulsive) vertex. Dashed blue line is the prediction in the strong coupling limit (eq. S4.12).

## S6. Detuning-dependence of the pairing energy

In figure 3a-c of the main text we showed three charge stability diagrams measured for three different polarizer detunings, demonstrating that the attractive corrections to the interaction between the system electrons increase with increasing $\delta$. We then explained using Little's chain model why the binding energy gained by two electrons that become nearest neighbors should exactly equal the energy stored in the polarizer, $\delta$.

In this section we expand on this in three aspects: First, we show additional data points, spanning more systematically the detuning axis and demonstrating more quantitatively the detuning dependence over the entire range of detunings. Second, we demonstrate how the intuition gained from the chain toy model in the main text is similarly valid even when we consider two sites, as in our experiments. Lastly, we show that there is a particle-hole symmetry in our experiments around $\delta = U$ and explain how this reflects a symmetric process of dressing holes by the polarizer.

**S4.1 Additional Data**

Figure S3 summarizes the measured dependence of $2\Delta$ on $\delta$, where for each point in this graph we set $\delta$, measure a charge stability diagram similar to those in Figs. 3a-c in the main text, and extract $2\Delta$ from the magnitude of the horizontal interaction vertex in this diagram. Notably, for small values of $\delta$, $2\Delta$ increases linearly with $\delta$. Around $\delta = 0$ the system electrons are repulsive ($2\Delta < 0$), but at a finite value of $\delta$, $2\Delta$ becomes positive and the electrons become attractive. The linear dependence fits rather well a line with unit slope ($2\Delta = \delta - W$, dashed purple line, where $W$ is the bare repulsion at $\delta = 0$) as is expected in the strong coupling limit (eq. S4.11). One can also notice that around the finite detuning, $\delta \approx 2.4 meV \approx U$, this trend is reversed and the pairing energy starts decreasing with a further increase of $\delta$. The decrease has roughly the same unit slope but with opposite sign. Notably, the measured detuning dependence fits very well, over the entire range of detunings, the prediction of the strong coupling limit, which is described by eq. S4.10 and shown in the figure as a purple dashed line.

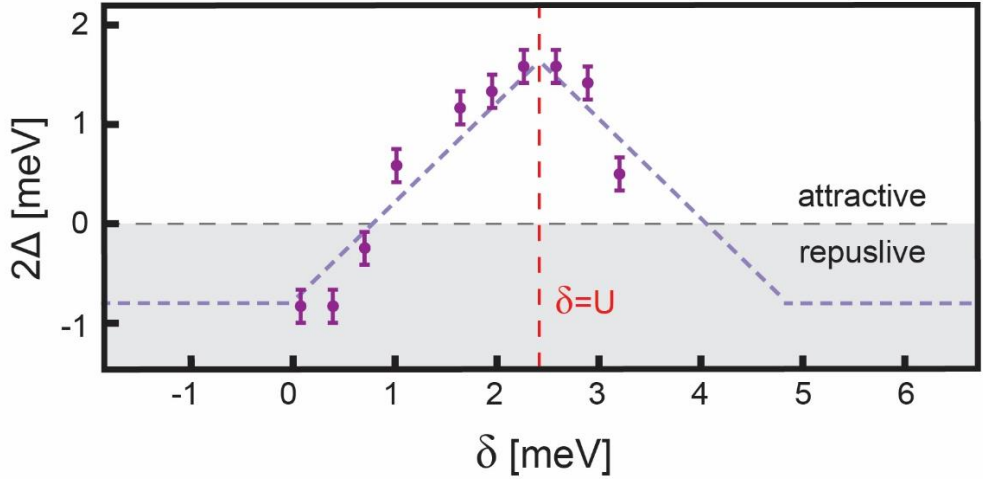

**Figure S7: Detuning dependence of the pairing energy.** For each point in the graph we set the detuning of the polarizer, $\delta$ (which is normalized to energy units using the measured lever-arm factor), measure a charge stability diagram similar to that in Fig. 2b in the main text, and extract $2\Delta$ from the length of the horizontal vertex. Negative values of $2\Delta$ correspond to the length of a vertical (repulsive) vertex. The dashed purple lines are the prediction in the strong coupling regime (eq. S4.10). They have slopes of $+1$ and $-1$, mirror reflected around the point $\delta = U$, (red dashed vertical line).

**S4.2 Binding energy equals the stored energy in the polarizer also for two sites**

In the main text we rationalized why the pairing energy is proportional to $\delta$ using Little's multi-site chain model. We explained that in the limit of strong charge-dipole coupling between an electron in the chain and the polarizers, an electron is dressed by the polarization of its nearby polarizers. The self-energy of this dressed particle thus includes the energy cost of polarizing the polarizers ($\delta$ per polarizer). We showed that two separate electrons would gain energy by pairing as nearest neighbors because then they share one polarizer and therefore share its polarization cost. The binding energy of this pair thus equals $\delta$, the stored energy in a single polarizer. In this section, we want to show in more quantitative detail that this simple picture of energy gain due to the sharing of polarization costs, obtained from the analysis on a chain, is also valid for two sites.

We start in fig. S8a with the simplest charge stability diagram: that of a bare system (without a polarizer) and with no repulsion between the electrons ($W = 0$). Note that the diagram is plotted in rotated axes with respect to the charge stability diagrams in the main text, namely, as a function of the potentials of the left and right wells in energy units, $v_L$ and $v_R$, and not of their difference and mean value as in figs. 2-4 in the main text. From the full Hamiltonian given in eq. S4.1, we can see that the charging of the left and right electrons occur when their chemical potentials equal zero ($\mu_L = 0$ and $\mu_R = 0$), namely, when the corresponding gate potentials equal their single-particle energy, $v_L = \epsilon_L$ and $v_R = \epsilon_R$. This picture changes considerably when we add a polarizer that is strongly coupled to the electrons (Fig. S8b). Consider for example the process of an electron entering the left site of the system. Due to the strong coupling with the polarizer, this electron cannot enter the system unless it is accompanied by the polarization of the polarizer. Consequently, what enters the system is an electron dressed by a polarization. This is clearly observed if we compare the two ground states before and after the charging (illustrations to the left and right of the bottom vertical charging line). The self-energy of this dressed particle now has to include the additional energy cost of the polarization, and thus the corresponding charging line shifts to $v_L = \epsilon_L + \delta$ (see figure). If, however, the left electron enters the system after the right electron has already entered, then it does not need to pay this polarization cost because the right electron has already triggered the polarization. The charging line that corresponds to this event (top vertical line) thus stays at the single particle value, $v_L = \epsilon_L$. The two electrons therefore share the polarization cost of the single polarizer. This is also understood by realizing that the chemical potential for the addition of two electrons at once, a process happening along the red diagonal line, follows the equation: $v_L + v_R = \left(\epsilon_L + \frac{\delta}{2}\right) + \left(\epsilon_R + \frac{\delta}{2}\right)$. Clearly, per electron the self-energy includes now only $\delta/2$, explaining why sharing the polarization costs makes the paired state more stable by $\delta$.

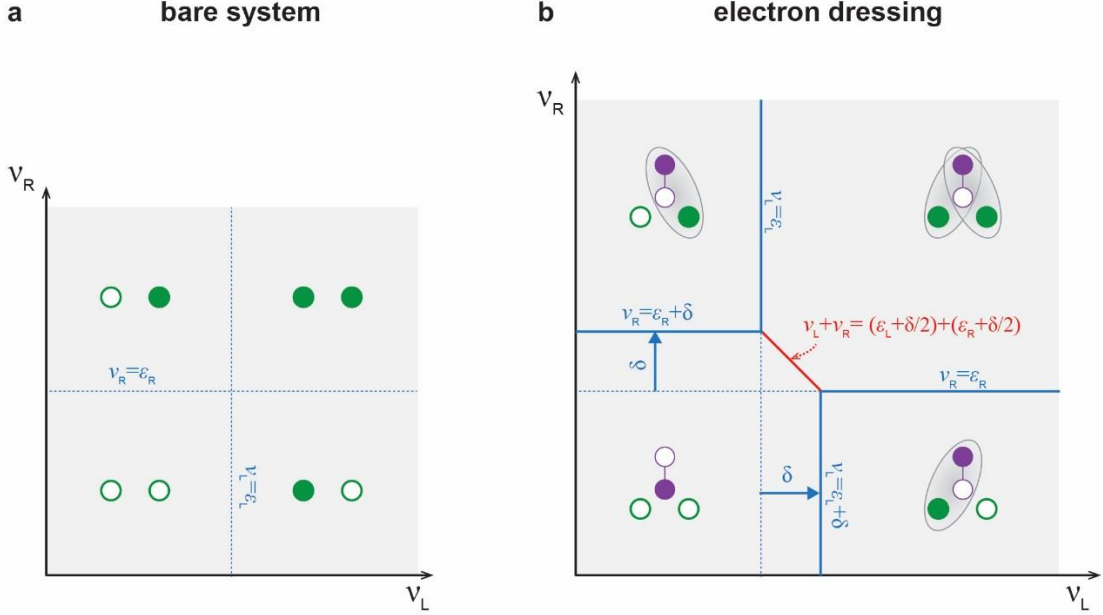

**Figure S8: Sharing the cost of polarization dressing by two electrons. a,** The charge stability diagram of the bare two-site system (without polarizer) and assuming for simplicity no repulsion between electrons. In contrast to the charge stability diagrams in the main text that are plotted as a function of the difference between the left and right well potentials and their mean value, here it is plotted as a function of the potentials themselves, $v_L$ and $v_R$. Full and empty circles correspond to a site occupied or unoccupied by a single electron. The charging lines of the left and right electrons happen at $v_L = \epsilon_L$ and $v_R = \epsilon_R$, marked by dotted blue lines. **b,** The charge stability diagram of the system with a polarizer in the strong coupling limit. System electrons are dressed by a polarization of the polarizer (marked by a gray ellipse). The charging lines of the dressed electrons (solid blue) are shifted by $\delta$ with respect to their non-interacting position (dotted blue). The charging line for populating two electrons simultaneously (red diagonal line) follows the formula $v_L + v_R = (\epsilon_L + \delta/2) + (\epsilon_R + \delta/2)$, reflecting the fact that both electrons share the polarization and thus each of them has only $\delta/2$ in its self-energy.

**S4.3 Particle-hole symmetry of the dressing by a polarizer**

As was shown in section S4.1, our measurements show that the detuning dependence of the pairing energy is symmetric around the point $\delta = U$. Below we show that this is a natural consequence of a particle-hole symmetry in our system. Note that this particle-hole symmetry should not be confused with the physical electron and hole states (above and below the dirac neutrality point), but is instead a symmetry within the band of electrons which maps any empty state to a occupied state and vice versa. Thus, under this particle hole symmetry the (0,0) state of the system (which can be considered as the bottom of the band) is mapped to the (1,1) state (which is mapped to the top of a one-electron-per site band). Similarly, under this symmetry the (1,0) state transforms to the (0,1) and vice versa, and the $|\Uparrow\rangle$ state of the polarizer is transformed to the $|\Downarrow\rangle$ state and vice versa.

When $\delta$ becomes larger than $U$, a single electron entering the system ceases to be dressed by a polarization, because the energy required for polarization is now larger than the energy gain of avoiding its repulsion from the electron in the polarizer. Instead, it is the holes that start getting dressed in a completely symmetric fashion to the dressing of the electrons for $\delta < U$. To see this, we should realize that the potentials as observed from the point of view of holes, $\bar{v}_L$ and $\bar{v}_R$, should be related to the potentials of the electrons, $v_L$ and $v_R$, using:

$$\bar{v}_L = U - v_L \quad \text{and} \quad \bar{v}_R = U - v_R$$

The minus sign in the equations above reflect the fact that the potential for holes is opposite to that for electrons, and the offset by $U$ results from the fact that for the holes the potential should be counted from the 'top of the band' rather than its bottom.

Fig. S9 shows the charge stability diagram for $\delta > U$, plotted as a function of the hole potentials, $\bar{v}_L$ and $\bar{v}_R$. Clearly, this charge stability diagram is completely symmetric to that of the electrons with $\delta < U$ (Fig. S8b). Specifically we see that in the top-right corner of the charge stability diagram there are no holes in the system and the polarizer has a hole on its bottom site (the symmetric case of a polarizer having an electron on its bottom site appearing in the bottom-left corner of the electrons' charge stability diagram, Fig. S8b). When one hole populates the right or left sites, it is accompanied by a polarization dressing (see illustrations). Correspondingly, the self-energy of the dressed hole is increased by $\delta$, causing a shift of its charging line by $\delta$, just as was seen for the electrons. When two holes populate the two sites they both share the polarization, each having only $\delta/2$ increase in their self-energy. The physics thus has a natural particle-hole symmetry built-in as is observed in the experiments.

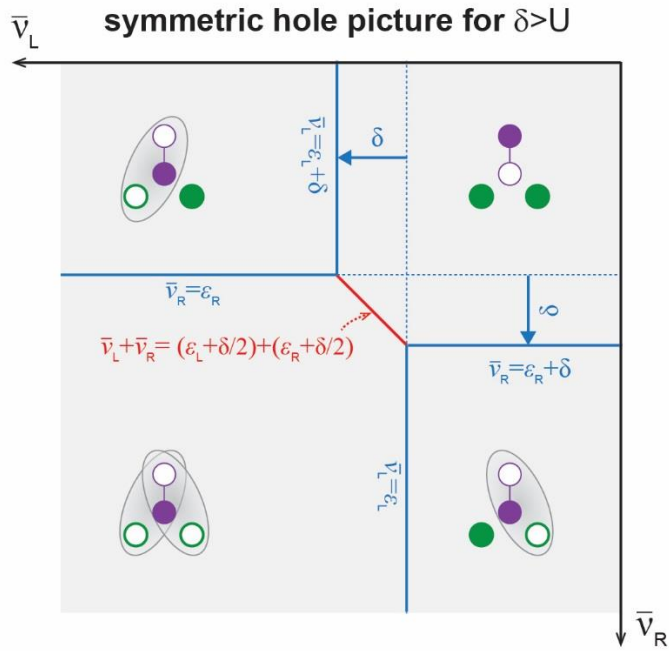

**Figure S9: Particle-hole symmetry of the dressing by a polarizer. a,** Charge stability diagram for $\delta > U$. The axes of this diagram are the hole voltages, $\bar{v}_L$ and $\bar{v}_R$, related to the electron voltages, $v_L$ and $v_R$, by $\bar{v}_L = U - v_L$ and $\bar{v}_R = U - v_R$ (see text). The top right corner now represents a state with zero holes and a polarizer with a hole at its bottom site, the completely symmetric counterpart of the electron state in the bottom left part of the charge stability diagram in fig. S8b. The gray ellipses now mark the dressing of holes by a polarization. Noticeably, the physics in the entire charge stability diagram is perfectly symmetric to that in fig. S8b.

## S7. Theoretical calculation of transport

In this section, we discuss the theoretical model used to calculate the transport through the polarizer-NT when it has strong correlations with the system-NT (Fig. 4 in the main text). As explained in the main text, to simplify the transport through the polarizer we opened the barrier between its top dot and the nearby lead making them strongly connected (Fig. 4a in the main text). As a result, the polarizer effectively became a single quantum dot coupled to two leads. Similar to the double quantum dot case, also a single quantum dot can act as a polarizer, only that now the polarization occurs between the dot and its lead instead of between the two dots. Although this single-dot polarizer configuration does not provide the simple fermion-parity-conserving polarization dressing that is provided by the double-dot polarizer (studied in the bulk of the paper) this configuration still allows us to study how the strong coupling between the system and polarizer electrons leads to highly-correlated transport processes. To keep the same notation as in the previous sections that discuss a double-dot polarizer we will label the two polarization states of the single dot (electron in the dot and electron in the lead) as $p = \Uparrow$ and $p = \Downarrow$ respectively. The transport process that would be of greatest interest to us, from the point of view of whether the pair that we created can coherently tunnel as a quantum mechanical entity, is the process involving the transition between the two even states, $|00, \Uparrow\rangle$ and $|11, \Downarrow\rangle$, a high-order process that involves the coherent tunneling of three electrons: a pair of electrons tunneling into the system and a single electron tunneling out of the polarizer and vice versa.

The current through the polarizer is calculated using a Master equation approach, where the transition rates are evaluated up to the lowest order in which pair tunneling can be observed. To capture the dynamics of the system, we add to the Hamiltonian in eq. S4.1 the tunnel couplings between the leads and the dots. We model the polarizer as a single dot connected to source ($S$) and drain ($D$) leads with tunneling $t_S$, $t_D$ respectively. The tunneling between the right (/left) system dot and its corresponding lead are described by $t_R$ (/$t_L$). The tunneling Hamiltonian is given by

(S7.1) $H_T = \sum_{\alpha \in \{S,D\}, k_\alpha} t_\alpha \left( \psi^\dagger_{k_\alpha} \psi_B + h.c. \right) + \sum_{k_L} t_L \left( \psi^\dagger_{k_L} \psi_L + h.c. \right) +$
$\sum_{k_R} t_R \left( \psi^\dagger_{k_R} \psi_R + h.c. \right)$

where the $k_\alpha$ indices denote the lead-states.

## Master equation approach

Within a master equation treatment of the QD dynamics, the occupation probabilities $P_m$ for the different states $m = |n_L\, n_R, p\rangle$ of the system + polarizer, are given by the rate equations

(S7.2) $\quad \dot{P}_m = -P_m \sum_{m' \neq m} \Gamma_{m \to m'} + \sum_{m' \neq m} P_{m'}\, \Gamma_{m' \to m}$.

Together with the normalization condition, $\sum_m P_m = 1$, the rate equations are solved for the steady-state occupation probabilities $P_m$, i.e. $\dot{P}_m = 0$. From the steady-state solution, the current through the polarizer dot can be obtained by evaluating the net rate of electrons flowing into lead $\alpha$ by flipping the polarizer between the $\Uparrow/\Downarrow$ states

(S7.3) $\quad I_\alpha = \mp e \sum_{n_L, n_R, n'_L, n'_R} \left( P_{n_L n_R, \Downarrow} \Gamma^\alpha_{n_L n_R, \Downarrow \to n'_L n'_R, \Uparrow} - P_{n_L n_R, \Uparrow} \Gamma^\alpha_{n_L n_R, \Uparrow \to n'_L n'_R, \Downarrow} \right)$

for $\alpha = S/D$, respectively.

## *T*-matrix evaluation of the transition rates

We use the generalized Fermi's golden rule for the transition rate between an initial $m$ and final $n$ state of the system+polarizer

(S7.4) $\quad \Gamma_{m \to n} = \frac{2\pi}{\hbar} \sum_{i'f'} P_{i'} |\langle f|T|i\rangle|^2\, \delta(E_f - E_i)$,

where the sum is over all possible initial $|i'\rangle$ and final $|f'\rangle$ states of the leads, $P_{i'}$ is the probability for the leads to be in the initial state $|i'\rangle$, $|i\rangle = |m\rangle \otimes |i'\rangle$, $|f\rangle = |n\rangle \otimes |f'\rangle$ and $T = H_T + H_T\, G_0\, H_T + H_T\, G_0\, H_T\, G_0\, H_T + \cdots$, $G_0 = \frac{1}{E_i - H_0}$, is the $T$-matrix. In the following we will discuss transitions between initial and final states which give rise to current through the polarizer, namely flip the polarizer state between $\Uparrow$ and $\Downarrow$.

1. 0e processes

To lowest order in the coupling between the dots and the leads, the transitions between states are governed by sequential tunneling processes. In these processes the charge state of the system doesn't change, and only the polarizer flips its polarization. The rates of these processes are given by Fermi's golden rule,

(S7.5) $\quad \Gamma^\alpha_{n_L n_R, \Downarrow \to n_L n_R, \Uparrow} = \Gamma_\alpha f_\alpha(E_{n_L n_R, \Uparrow} - E_{n_L n_R, \Downarrow})$

(S7.6) $\quad \Gamma^\alpha_{n_L n_R, \Uparrow \to n_L n_R, \Downarrow} = \Gamma_\alpha [1 - f_\alpha(E_{n_L n_R, \Downarrow} - E_{n_L n_R, \Uparrow})]$,

where $\Gamma_\alpha = 2\pi \rho_\alpha |t_\alpha|^2$, $\rho_\alpha$ is the density of states in lead $\alpha$, $f_\alpha$ is the Fermi-Dirac distribution of lead $\alpha$ and $\alpha$ denotes the lead through which the tunneling process occurs.

2. 1e processes

The next-to-leading order terms in the expansion of the $T$-matrix give rise to processes where an electron is exchanged between one of the dots and its lead, in addition to flipping the polarizer. An example for such a process is

(S7.7) $\quad |10, \Downarrow\rangle \leftrightarrow \begin{matrix} |00, \Downarrow\rangle \\ |10, \Uparrow\rangle \end{matrix} \leftrightarrow |00, \Uparrow\rangle$

where a similar process happens for an electron in the right dot. The transition rates for these processes are given by

(S7.8) $\quad \Gamma^\alpha_{10\Downarrow,00\Uparrow} = \frac{2\pi}{\hbar} \sum_{\alpha i_\alpha f_\alpha} P_{i_\alpha} |\langle f_\alpha|\langle 00, \Uparrow|H_T G_0 H_T|10, \Downarrow\rangle |i_\alpha\rangle|^2 \delta(E_f - E_i)$

$= \frac{\Gamma_\alpha \Gamma_L}{2\pi \hbar} \int d\varepsilon_{k_L} \int d\varepsilon_{k_\alpha} \left| \sum_{i_1} \frac{1}{E_i - E_{i_L} \pm \varepsilon_{k_{i_L}}} \right|^2 f_\alpha(\varepsilon_{k_\alpha})[1 - f_L(\varepsilon_{k_L})]\delta(\Delta E_{10\Downarrow,00\Uparrow} + \varepsilon_{k_L} - \varepsilon_{k_\alpha})$

$= \frac{\Gamma_\alpha \Gamma_L}{2\pi \hbar} \int d\varepsilon_{k_L} \left| \frac{1}{\varepsilon_{k_L} - \Delta E_{00\Downarrow,10\Downarrow}} - \frac{1}{\varepsilon_{k_L} + \Delta E_{10\Uparrow,00\Uparrow} + \Delta E_{10\Downarrow,00\Uparrow}} \right|^2 f_\alpha(\Delta E_{10\Downarrow,00\Uparrow} + \varepsilon_{k_L})[1 - f_L(\varepsilon_{k_L})]$

where $\alpha \in \{S, D\}$, $+/-$ in the denominator in the second line is for intermediate states with an electron added/removed to/from the dots, and the intermediate lead state is $k_{i_L} \in \{k_L, k_\alpha\}$.

For the reverse process we have

(S7.9) $\quad \Gamma^\alpha_{00\Uparrow,10\Downarrow} = \frac{\Gamma_\alpha \Gamma_L}{2\pi \hbar} \int d\varepsilon_{k_1} \left| \frac{1}{\varepsilon_{k_L} - \Delta E_{00\Uparrow,10\Downarrow} - \Delta E_{00\Downarrow,00\Uparrow}} - \frac{1}{\varepsilon_{k_L} + \Delta E_{10\Uparrow,00\Uparrow}} \right|^2 f_L(\varepsilon_{k_L})[1 - f_\alpha(\varepsilon_{k_L} - \Delta E_{00\Uparrow,10\Downarrow})]$.

The rates of the 1e processes can be evaluated analytically following the standard procedure for cotunneling.

3. 2e processes

Here, we are interested in the higher-order tunneling process (third term in the expansion of the $T$-matrix) which allows for direct pair tunneling between the even states of the system, $|00,\Uparrow\rangle$ and $|11,\Downarrow\rangle$. There are six different intermediate pathways between the initial and final states

(S7.10) $\quad |11,\Downarrow\rangle \leftrightarrow \begin{matrix} |11,\Uparrow\rangle \leftrightarrow |01,\Uparrow\rangle \\ |11,\Uparrow\rangle \leftrightarrow |10,\Uparrow\rangle \\ |10,\Downarrow\rangle \leftrightarrow |00,\Downarrow\rangle \\ |10,\Downarrow\rangle \leftrightarrow |10,\Uparrow\rangle \\ |01,\Downarrow\rangle \leftrightarrow |00,\Downarrow\rangle \\ |01,\Downarrow\rangle \leftrightarrow |01,\Uparrow\rangle \end{matrix} \leftrightarrow |00,\Uparrow\rangle.$

The transition rates are given by

(S7.11) $\quad \Gamma^{\alpha}_{11\Downarrow,00\Uparrow} = \frac{2\pi}{\hbar} \sum_{\alpha i_\alpha f_\alpha} P_{i_\alpha} |\langle f_\alpha | \langle 00,\Uparrow | H_T G_0 H_T G_0 H_T | 11,\Downarrow\rangle |i_\alpha\rangle|^2 \delta(E_f - E_i)$

$= \frac{\Gamma_\alpha \Gamma_L \Gamma_R}{(2\pi)^2 \hbar} \int d\varepsilon_{k_L} \int d\varepsilon_{k_R} \int d\varepsilon_{k_\alpha} \left| \sum_{i_L i_R} \frac{1}{\left(E_i - E_{i_R} \pm \varepsilon_{k_{i_R}} \pm \varepsilon_{k_{i_L}}\right)\left(E_i - E_{i_L} \pm \varepsilon_{k_{i_L}}\right)} \right|^2 \times$

$f_\alpha(\varepsilon_{k_\alpha})[1 - f_L(\varepsilon_{k_L})][1 - f_R(\varepsilon_{k_R})]\delta(\Delta E_{11\Downarrow,00\Uparrow} + \varepsilon_{k_L} + \varepsilon_{k_R} - \varepsilon_{k_\alpha}) =$

$\frac{\Gamma_\alpha \Gamma_L \Gamma_R}{(2\pi)^2 \hbar} \int \varepsilon_{k_L} \int d\varepsilon_{k_R} \left| \sum_{i_L i_R} \frac{1}{\left(E_i - E_{i_R} \pm \varepsilon_{k_{i_R}} \pm \varepsilon_{k_{i_L}}\right)\left(E_i - E_{i_L} \pm \varepsilon_{k_{i_L}}\right)} \right|^2 \times f_\alpha(\varepsilon_{k_L} + \varepsilon_{k_R} +$

$\Delta E_{11\Downarrow,00\Uparrow})[1 - f_L(\varepsilon_{k_L})][1 - f_R(\varepsilon_{k_R})],$

and

(S7.12) $\quad \Gamma^{\alpha}_{00\Uparrow,11\Downarrow} = \frac{\Gamma_\alpha \Gamma_L \Gamma_R}{(2\pi)^2 \hbar} \int d\varepsilon_{k_L} \int d\varepsilon_{k_R} \left| \sum_{i_L i_R} \frac{1}{\left(E_i - E_{i_R} \pm \varepsilon_{k_{i_R}} \pm \varepsilon_{k_{i_L}}\right)\left(E_i - E_{i_L} \pm \varepsilon_{k_{i_L}}\right)} \right|^2 \times$

$f_L(\varepsilon_{k_L}) f_R(\varepsilon_{k_R})[1 - f_\alpha(\varepsilon_{k_L} + \varepsilon_{k_R} - \Delta E_{00\Uparrow,11\Downarrow})],$

respectively, where $+/-$ is for intermediate states with an electron added/removed to/from the QD system, $k_{i_{L/R}} \in \{k_L, k_R, k_\alpha\}$, and $\Delta E_{11\Downarrow,00\Uparrow} = -\Delta E_{00\Uparrow,11\Downarrow} = \varepsilon_d - (\varepsilon_L + \varepsilon_R + U)$.

Evaluating the rates at finite $T$ in the polarizer

Assuming zero temperature in the leads coupled to the system and finite temperature in the $S/D$ leads coupled to the polarizer, the rates can be calculated analytically in the low-bias limit where the energy dependence from the intermediate states in the denominators can be neglected.

For the $|11,\Downarrow\rangle \to |00,\Uparrow\rangle$ process, we then find

(S7.13) $\int_{\mu_L}^{\infty} d\varepsilon_{k_L} \int_{\mu_R}^{\infty} d\varepsilon_{k_R} f_\alpha(\varepsilon_{k_L} + \varepsilon_{k_R} + \Delta E_{11\Downarrow,00\Uparrow}) = -(k_B T)^2 \int_a^{\infty} dx [x - \ln(1 + e^x)] = (k_B T)^2 \left[\frac{1}{2} a^2 + \frac{\pi^2}{6} + \text{Li}_2(-e^a)\right]$

where $a = (\mu_L + \mu_R + \Delta E_{11\Downarrow,00\Uparrow} - \mu_\alpha)/k_B T$ and $\text{Li}_2(z)$ is the polylogarithm function. For the opposite process $|00, \Uparrow\rangle \to |11, \Downarrow\rangle$, we use the identity $1 - f(\varepsilon) = f(-\varepsilon)$ and reach the same result with $a = (\Delta E_{00\Uparrow,11\Downarrow} + \mu_\alpha - \mu_1 - \mu_2)/k_B$.

## S8. Supplementary Discussion

In this paper we demonstrated 'excitonic' attraction between electrons using the basic building block that exhibits this physics. These results raise a set of important questions on whether it would be possible to engineer interesting states of matter by generalizing this single block to multiple blocks. For example: Would it possible to create an artificial superconductor in this way? Would such a superconductor be stable against competing ground states? What would the nature of electron pairing be in such a superconductor? Can the pairing energy be pushed to even higher values than those measured in the current work? In this section we briefly address these different aspects.

Before discussing these questions, we would like to mention an alternative mechanism that was discussed in the literature for electron pairing, driven only by disorder. As shown by Shepelyansky[6], in the presence of disorder two repelling electrons can coherently propagate together over longer distances than is naively expected. Unlike Little's model, this mechanism does not require a surrounding medium with which the electrons strongly interact. However, this mechanism applies only to highly excited states and not the ground state and therefore is not relevant for realizing an engineered superconductor, which is the aim of Little's proposal. We therefore do not discuss this mechanism further, and return to analyze the possibility of realizing a macroscopic superconductor using the basic building block presented in our work.

The most natural generalization of the building block that we have demonstrated is to the one-dimensional 'molecule' that was originally proposed by Little[5]. In this molecule the system comprises multiple sites along a linear chain and has a polarizer located between each pair of sites (Fig. S10a). Theoretically this model has been studied extensively in the literature, especially in the context of whether such a molecule could superconduct and what are the optimal parameters for getting superconductivity with the largest critical temperature. An important point that was realized already early on was that the absence of long range order in one-dimension prohibits a true superconducting ground state in this molecule in the thermodynamic limit, but does allow for superconducting correlations over finite lengths. The original work by Little[5] analyzed this model under very specific conditions, assuming that the polarizers' response is retarded, similar to the behavior of phonons in the BCS theory. Little's model has however other physical regimes that might be even more relevant for achieving superconductivity. Specifically the polarizer's response can be instantaneous or retarded, and their coupling to the electrons in the system can be in the weak or strong limit (defined quantitatively below). An exhaustive analysis of this model, in all these regimes, was done by Hirsch and Scalapino[4], in an attempt to test whether there is a parameter range in which superconducting correlations dominate over the two relevant competing orders – charge and spin density waves. Specifically, they calculated the dominant order for various interaction strengths $U$, $W$ and tunneling rate in the polarizer, $t$, and between system sites, $t_s$. These calculations showed that in the original regime considered by Little, superconductivity is not the dominant order, as well as in many other of the regimes in parameters space. They did demonstrate, however, that superconducting correlations are clearly the dominant order in a certain range in parameter space, where the interaction with the dipole is instantaneous ($t > t_s$) and the charge dipole coupling is in the strong limit ($U > t$). Interestingly, our experiments demonstrate the pairing exactly in this regime. The experiments in Figs. 2b, 3a-c and 4 are done in the strong coupling limit (in fig. 4, for example, $\Gamma_{pol}/U \approx 0.2$, where $\Gamma_{pol}$ is the tunneling rate to the polarizer). Moreover we observe clear pairing over the entire range going from the instantaneous limit shown in Figs. 2 and 3 of the main text to the retarded limit observed in Fig. 4 (for which $\Gamma_{pol}/\Gamma_R = 0.9$ and $\Gamma_{pol}/\Gamma_L = 0.44$, where $\Gamma_L$, $\Gamma_R$ are the tunneling rates into the left and right dots, namely, the polarizer is slower than the system).

There is one important difference between Little's model and the building block in our experiment: Little considered spinful electrons and in his model the pairing, induced by the polarizers, occurs on a single system site between two electrons of opposite spin. The pairing energy gain thus has to compete with the on-site repulsion. In contrast, in our experiment (and similarly in the theory by Raikh, Glazman and Zhukov who first considered the basic building block theoretically[7]) the only relevant degree of freedom is that of the charge and the pairing occurs between nearest neighbors. In this case the attraction does not need to compete with the onsite repulsion, which is often very large, but rather with the nearest neighbor repulsion. The advantage of this is that it allows to create much stronger attraction, even reaching the instantaneous limit. One question that may arise, though, is whether nearest neighbor attraction on a one-dimensional chain (Fig. S10a) will not lead to phase separation, namely to all the electrons bunching into a macroscopic clump of charge rather than to the formation of pairs. Theoretically, the question of whether the paired or phase-separated state would be more stable, especially in the limit in which the tunneling along the system is comparable to the attraction, is still not known. There is however, a simple variation of our model that resolve this problem while maintaining the other benefit: Consider, that our basic units are organized in a "ladder" type structure (Fig. S10b), such that our original left/right sites appear now at the two sides of a single rung. In this case the pair would form along a rung, and breaking it by separating the two electrons to different rungs will cost energy, since one more polarizer will be needed to be polarized, again giving the pairing energy. Note that this model is completely equivalent to Little's model, only that instead of the spin degree of freedom we now have an 'isospin' degree of freedom describing whether an electron is on the left or right leg of the ladder. The ladder realization is accurately described by a negative $U$ hubbard model, which at least in two dimensions (2D) is known to support superconductivity[8].

We note that the building block demonstrate here is not restricted to 1D and can be also implemented with quantum dots in 2D, using either the highly-advanced techniques for making quantum dot arrays in semiconductors[9–12], or even using layered van-der-Waals materials[13] to create a 2D electronic system adjacent to double-layer polarizable stack. An important feature of attraction that is driven only by Coulomb repulsion that it scales like

Coulomb law, inversely proportional to the dimensions of the underlying components. If the building block that we demonstrated here can be scaled down to a $\sim 1nm$ scale, for example, by making a 2D array of electronically clean quantum dots by arranging atoms on the surface of an insulator or a semiconductor, as has been demonstrated experimentally recently[14], one could expect the pairing energy observed here to grow much beyond room temperature.

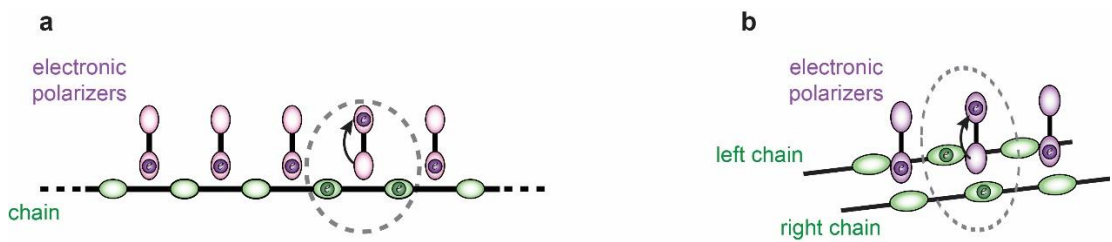

**Figure S10. Possible scaling up the excitonic pairing building block in one-dimension**. **a,** The "Little chain". The system sites are arranged along a 1D chain, with a polarizer between each two of sites. The electron pairs travel along the main system chain. **b,** An alternative 'ladder' design, in which the system comprises two 1D chains and the polarizers are in between. Here the pairing occur along the rungs of the ladder.